\documentclass[nofootinbib,aps,preprint]{revtex4}

\usepackage{amsmath} 
\usepackage{epsfig}
\usepackage{cancel}
\usepackage{color}
\usepackage{amssymb}
\usepackage{pstool}
\usepackage{pstricks}
\usepackage{graphicx}
\usepackage{psfrag}
\usepackage{comment}

\newcommand{\D}{D^{0}}
\newcommand{\Dbar}{\overline{D}^{0}}
\newcommand{\K}{K^{0}}
\newcommand{\Kbar}{\overline{K}^{0}}

\newcommand{\ket}[1]{|#1\rangle}
\newcommand{\bra}[1]{\langle#1|}

\newcommand{\Acal}{\mathcal{A}}

\newcommand{\real}{\textrm{Re}}
\newcommand{\imag}{\textrm{Im}}

\usepackage{xcolor}
\definecolor{red}{rgb}{1.0, 0, 0}
\definecolor{blue}{rgb}{0, 0, 1.0}

\def\beq{\begin{equation}}
\def\eeq{\end{equation}}
\def\beqa{\begin{eqnarray}}
\def\eeqa{\end{eqnarray}}
\def\ben{\begin{enumerate}}
\def\een{\end{enumerate}}

\begin{document}  
  
  
\vspace*{18pt}   
  
\title{\boldmath Effects of $\K-\Kbar$ mixing on determining $\gamma$ from $B^{\pm}\to DK^{\pm}$}

\author{Yuval Grossman}     
\author{Michael Savastio}
\affiliation{Department of Physics, LEPP, Cornell University, Ithaca, NY 14853 \vspace*{8pt}}
  
\begin{abstract} \vspace*{10pt} 
The decay $B^{\pm}\to DK^{\pm}$ followed by the subsequent decay of the $D$ meson into final states involving a neutral kaon can be used to determine the CKM angle $\gamma$.  
We study CP violation effects due to mixing and decay of the final state kaon. We find that ignoring these effects produces a shift in $\gamma$ of order $\epsilon_{K}/r_{B}$, an enhancement of $1/r_B$ compared to the naive expectation. 
We then show how to take these effects into account such that, in principle, they will not introduce any theoretical error in the extraction of $\gamma$.
\end{abstract}  
  
\maketitle

\section{Introduction}

Interference between the $b\to c\bar{u}s$ and $b\to u\bar{c}s$ decay amplitudes can be used to determine the weak phase
\begin{equation}
\gamma\equiv\arg\left(-\frac{V_{ud}^{\phantom{*}}V_{ub}^{*}}{V_{cd}^{\phantom{*}}V_{cb}^{*}}\right)\,.
\end{equation}
There are many hadronic final states that can be used to add information about $\gamma$~\cite{BLS,Gronau:1991dp,Atwood:2000ck,Giri:2003ty,Poluektov:2004mf,Gronau:1990ra,Kayser:1999bu} and, what is relevant to our case, some of them have one (or more) neutral kaon in the final state.  All methods of determining $\gamma$ from $B^{\pm}\to DK^{\pm}$ involve deriving a system of equations for $\gamma$ in terms of the decay widths and amplitudes for the various processes involved.  If all other quantities can be determined experimentally this allows for a model-independent determination of $\gamma$.

Thus far CP violation associated with neutral kaons in the
determination of $\gamma$ has been neglected since the effect is
expected to be much smaller than current experimental
uncertainties. As more statistics become available new sources of CP
violation in the overall process will become significant and will need
to be taken into account. In this work we study the effects of CP
violation in the kaon system
on the determination of $\gamma$. Such effects were studied in other systems
in \cite{Azimov:1980,Azimov:1998sz,Lipkin:1999qz,Bigi:2005ts,Calderon:2007rg,Grossman:2011zk}.
As we will see, there are effects
that are linear in $\epsilon_{K}$, and they are parametrically
enhanced by $1/r_B$ (we use standard notations that are defined
below). Once CP violation in the kaon system is included it is
crucial that the time dependence of the kaon mixing be taken into
account.  This must be done in a way which considers time dependent
detector efficiencies, and thus it can only be carried out by each experiment
separately as part of their analysis.

\section{GLW}

To understand the main issues associated with including kaon mixing and CPV effects, we start with the theoretically simplest case, which is the GLW method \cite{Gronau:1991dp,BLS}. We define
\begin{equation}
A(B^{-}\to\D K^{-})=A(B^{+}\to\Dbar K^{+})\equiv A_{B},
\label{def_AB}
\end{equation}
\begin{equation}
A(B^{-}\to\Dbar K^{-})\equiv A_{B}r_{B}e^{i(\delta_{B}-\gamma)}, \qquad A(B^{+}\to\D K^{+})\equiv A_{B}r_{B}e^{i(\delta_{B}+\gamma)},
\label{def_rB}
\end{equation}
where $\delta_B$ is a strong phase and $r_{B}$ is a real parameter
which accounts for color and CKM suppression, and is measured to be of
order $10^{-1}$. We consider $D \to K\pi^{0}$ decays, but the
$\pi^{0}$ can be exchanged for another CP eigenstate such as
$\rho^{0}$ and the same discussion would apply. We further assume
\begin{equation}
|A(\D \to \Kbar \pi^{0})| = |A(\Dbar \to \K \pi^{0})|, \qquad
|A(\D \to \K \pi^{0})| =|A(\Dbar \to \Kbar \pi^{0})| = 0.
\end{equation}
The decay $A(\D\to\K\pi^{0})$ and its CP conjugate are doubly
Cabbibo suppressed, and there is no real complication in including them,  they are unimportant for our current discussion, so for the moment we set them to zero (they are included in our final results).  We also neglect terms of order $r_{B}|\epsilon_{K}|$ in this section for simplicity.  For the kaons we use the standard notation
\begin{equation} \label{eq:kaondef}
\ket{\K}=\frac{1}{2p}\left(\ket{K_L}+\ket{K_S}\right),\qquad
\ket{\Kbar}=\frac{1}{2q}\left(\ket{K_L}-\ket{K_S}\right),\qquad
\frac{A(K_L \to \pi \pi)}{A(K_S \to \pi \pi)}
=\epsilon,
\end{equation}
where in the last term we neglected direct CPV in kaon decays, that
is, we set $\epsilon'=0$.
We define the time dependent asymmetry
\begin{equation}
a_{CP}(t)=\frac{\Gamma(B^+ \to (K_S \pi^0)_D K^{+})-\Gamma(B^- \to (K_S \pi^0)_D K^-)}
{\Gamma(B^+ \to (K_S \pi^0)_D K^+)+\Gamma(B^- \to (K_S \pi^0)_D K^-)},
\end{equation}
In the above $t$ is the proper time of the kaon system, and by
$K_S$ we refer to a kaon that decays into two pions (see discussion in
\cite{Grossman:2011zk}).
We work to first order in $\epsilon_{K}$ and we neglect terms
of order $r_{B}|\epsilon_{K}|$ to find
\begin{equation}
a_{CP}(t)=\frac{2r_{B}\sin(\gamma)\sin(\delta_{B})-2\real(\epsilon)+2e^{(\Gamma_{S}-\Gamma)t}\real(\epsilon^{*}e^{ix\Gamma t})}{1+r_{B}^{2}-2r_{B}\cos(\gamma)\cos(\delta_{B})}.
\label{aCP_timedependent}
\end{equation}
where as usual
\begin{equation}
\Gamma=\frac{\Gamma_{S}+\Gamma_{L}}{2}, \qquad x=\frac{m_{L}-m_{S}}{\Gamma}\,.
\end{equation}

It is useful to consider also the case where we integrate over the kaon lifetime. 
Following~\cite{Grossman:2011zk} we parametrize
the experiment-dependent efficiency to detect the kaon by 
$F(t)$ with $0 \le F(t) \le 1$. We emphasize that $F$ must
be determined as part of the experimental analysis. We then define the time integrated asymmetry
\begin{equation}
A_{CP}=\frac{\int F(t) \,dt\, [\Gamma(B^+ \to (K_S \pi^0)_D K^{+})-\Gamma(B^- \to (K_S \pi^0)_D K^-)]}
{\int F(t) \,dt\, [\Gamma(B^+ \to (K_S \pi^0)_D K^+)+\Gamma(B^- \to (K_S \pi^0)_D K^-)]},
\end{equation}
where the integral is from zero to infinity. To demonstrate the effect we take a simple case where $F(t)=1$ and we obtain
\begin{equation}
A_{CP}=
\frac{2r_{B}\sin(\gamma)\sin(\delta_{B})+2\real(\epsilon)}
{1+r_{B}^{2}-2r_{B}\cos(\gamma)\cos(\delta_{B})}.
\label{aCP}
\end{equation}
We can check few limits of the above results:
\begin{enumerate}
\item
For the case of no CPV in kaons, that is, $\epsilon\to0$, and to first
order in $r_B$ we obtain 
\beq
A_{CP}= 2r_{B}\sin(\gamma)\sin(\delta_{B}),
\eeq
as we should.
\item
The $r_{B}\to0$ limit corresponds to the case where the only source of
CP violation is in the kaon system.  This case was studied for
$\tau$ decays in \cite{Grossman:2011zk}. Using Eq.~(\ref{aCP}) for
$r_{B}\to0$ we see that 
\beq
A_{CP}=2\real(\epsilon),
\eeq 
in agreement with \cite{Grossman:2011zk}. 
\item
Last we work to first order in $r_B$ and we get
\begin{equation} \label{mainGLW}
A_{CP}=2r_B\left[\sin(\gamma)\sin(\delta_{B})+\frac{\real(\epsilon)}{r_B}\right].
\end{equation}
Therefore, when both effects are included, $A_{CP}$ (and therefore the
extracted value of $\gamma$) is shifted by a term of order $\real(\epsilon)/r_{B}$. It is the $1/r_B$ enhancement which makes the effect larger and somewhere at the level that is expected to be probed in the near future.
\end{enumerate}

\section{The general case: Dalitz decays}

\subsection{General discussion}

Our goal is to obtain a system of equations for $\gamma$ in terms of
various experimentally determined quantities in the most general case of a multi-body $D$ decay such as $D\to K_{S}\,\pi^{+}\pi^{-}$. The quantities involved will be integrated over finite regions of the $D$ decay phase space. 
Here we are using $D\to K_S\pi^{+}\pi^{-}$ for concreteness, however our
discussion will also apply to other decay modes such as $D\to
KK^{+}K^{-}$ and also to two-body decays, such as $D\to K_{S}\pi^{0}$, in which these amplitudes are not momentum dependent. We do not include other small effects like 
$\D-\Dbar$ mixing which was discussed in
\cite{Giri:2003ty,Grossman:2005rp,Silva:1999bd,Amorim:1998pi,Rama:2013voa}. Note that, though we will not discuss it further, it is imperative to include the effects of $\D-\Dbar$ mixing since they will be competitive with those of CP violation in $\K-\Kbar$ mixing.

With this in mind, we begin by defining quantities associated with the three-body $D$ decay
\begin{equation}
\Acal(t)\equiv A(\D\to(\pi\pi)_{K}\pi^{+}\pi^{-})=A_{\pi\pi}\left(A_{S}e^{-im_{S}t-\Gamma_{S}t/2}+\epsilon A_{L}e^{-im_{L}t-\Gamma_{L}t/2}\right),
\label{Acal}
\end{equation}
and its CP conjugate
\begin{equation}
\bar{\Acal}(t)\equiv A(\Dbar\to(\pi\pi)_{K}\pi^{-}\pi^{+})=A_{\pi\pi}\left(\bar{A}_{S}e^{-im_{S}t-\Gamma_{S}t/2}+\epsilon \bar{A}_{L}e^{-im_{L}t-\Gamma_{L}t/2}\right),
\end{equation}
where
\begin{equation}
A_{\pi\pi}\equiv A(K_{S}\to\pi\pi),\qquad A_{S,L}\equiv A(\D\to K_{S,L}\pi^{+}\pi^{-}),\qquad 
\bar{A}_{S,L}\equiv A(\Dbar\to K_{S,L}\pi^{-}\pi^{+}),
\end{equation}
and $\pi\pi$ is either of $\pi^{+}\pi^{-}$ or $\pi^{0}\pi^{0}$. 
Note that $\Acal$ and $\bar{\Acal}$ are analogous to $A_{D}(s_{12},s_{13})$ and $A_{D}(s_{13},s_{12})$ of \cite{Giri:2003ty} respectively.
We are assuming that we use the $\pi\pi$ final state to tag the kaon as a $K_{S}$ and
we have used (\ref{eq:kaondef}). From now on meson variables such as $p,q,\epsilon,x,y$ will be for the $K$ unless stated otherwise. 

Interference of the $\Acal$ and $\bar{\Acal}$ amplitudes will occur through the $B^{\pm}$ decay, where the relative phase of the interference involves $\gamma$.  With the definitions (\ref{def_AB}) and (\ref{def_rB}) we can write down the amplitude for the overall process in terms of $\gamma$
\begin{equation} \label{eq:deff}
A(B^{-}\to f^-)=A_{B}\left[\Acal(t)+r_{B}e^{i(\delta_{B}-\gamma)}\bar{\Acal}(t)\right], \qquad
f^\pm=[(\pi\pi)_{K}\pi^{+}\pi^{-}]_{D}K^{\pm}.
\end{equation}
One can obtain the amplitude for the $B^{+}\to f^+$ by  $\gamma\to-\gamma$ and $\Acal\leftrightarrow\bar{\Acal}$.  Squaring this we can write the time-dependent differential normalized widths
\begin{eqnarray} \label{mastercp}
d\hat\Gamma(B^{-}\to f^-)&=&|\Acal|^{2}+r_{B}^{2}|\bar{\Acal}|^{2}+2r_{B}\left[\real(\Acal^{*}\bar{\Acal})\cos(\delta_{B}-\gamma)-\imag(\Acal^{*}\bar{\Acal})\sin(\delta_{B}-\gamma)\right], \nonumber \\
d\hat\Gamma(B^{+}\to f^+)&=&|\bar{\Acal}|^{2}+r_{B}^{2}|\Acal|^{2}+2r_{B}\left[\real(\Acal^{*}\bar{\Acal})\cos(\delta_{B}+\gamma)+\imag(\Acal^{*}\bar{\Acal})\sin(\delta_{B}+\gamma)\right],~~~~~
\end{eqnarray}
where $\hat\Gamma \equiv \Gamma/{|A_{B}|^{2}}$.

Eqs.~(\ref{mastercp}) are our system of equations which can be used to
determine $\gamma$.  Note that $|\Acal|^{2}$ and $|\bar{\Acal}|^{2}$
are directly measurable in $D$ decays. In addition to these there is a phase, that of $\Acal^{*}\bar{\Acal}$, which is momentum dependent in the multi-body case and which must be obtained in addition to $r_{B}$ and $\delta_{B}$ in order to determine $\gamma$. These equations are identical in form to those from \cite{Giri:2003ty}, and similar to those in \cite{Gronau:1991dp} except that we distinguish between $\Acal$ and $\bar{\Acal}$.  The new complications are hidden in the time-dependence and CP violating parameters stored in $\Acal$  and $\bar{\Acal}$.

We should proceed by setting up equations for the Dalitz
analysis. This is performed by integrating over bins in the $D\to
K_S\pi\pi$ phase space.  First, let us define the momenta of the decay
products and the Mandelstam variables as in \cite{Giri:2003ty} 
\beq
K(p_{1}),\quad \pi^{-}(p_{2}),\quad
\pi^{+}(p_{3}), \qquad s_{ij}=(p_{i}+p_{j})^{2}.
\eeq
In the $\epsilon\to0$ limit one has
$\Acal(s_{12},s_{13})=\bar{\Acal}(s_{13},s_{12})$.  One way to think
of this is that if the $K_{S}$ were a CP eigenstate then
$K_{S}\pi^{+}\pi^{-}$ is a CP eigenstate but for momentum
interchange. The effect due to the fact that the $K_{S}$ is not a CP eigenstate is that $\Acal(s_{12},s_{13})=\bar{\Acal}(s_{13},s_{12})+O(|\epsilon|)$. 

We would like to partition the $(s_{12},s_{13})$ phase space into bins which are symmetric about the line $s_{12}=s_{13}$.  With $\epsilon\to0$ very simple relations exist between the various integrals above and below the $s_{12}=s_{13}$ line. The CP violation in the kaon system, however, complicates this. To this end, we define integrals over bins in phase space
\begin{equation}
T^{-}_{i}(t)\equiv\int_{i}d^{2}s\:|\Acal(t)|^{2}, \qquad T^{+}_{i}(t)\equiv\int_{i}d^{2}s\:|\bar{\Acal}(t)|^{2},
\end{equation}
\begin{equation}
C_{i}(t)\equiv\int_{i}d^{2}s\:\real(\Acal^{*}\bar{\Acal}),\qquad S_{i}(t)\equiv\int_{i}d^{2}s\:\imag(\Acal^{*}\bar{\Acal}),
\end{equation}
where we have emphasized that these quantities all depend on the proper time of the kaon. The index $i$ labels a bin in $(s_{12},s_{13})$ phase space, and we will let $\bar{i}$ denote the bin obtained by reflecting the bin $i$ about $s_{12}=s_{13}$.  We can then write a system of equations for integrals of the overall decay width in these bins. At this stage we assume that we integrate over time, and we obtain
\begin{equation}
\hat{\Gamma}^{-}_{i}=T^{-}_{i}+r_{B}^{2}T^{+}_{i}+2r_{B}\left[C_{i}\cos(\delta_{B}-\gamma)-S_{i}\sin(\delta_{B}-\gamma)\right],
\end{equation}
\begin{equation}
\hat{\Gamma}^{+}_{i}=T^{+}_{i}+r_{B}^{2}T^{-}_{i}+2r_{B}\left[C_{i}\cos(\delta_{B}+\gamma)+S_{i}\sin(\delta_{B}+\gamma)\right].
\end{equation}
Note that the quantities $T_{i}^{\pm}$ are the widths for $D\to K_{S}\pi^{+}\pi^{-}$ with the $K_{S}$ identified through its decay and thus can be determined from charm data, so we will treat these as known.  In the $\epsilon\to0$ limit we have $T_{\bar{i}}^{-}=T^{+}_{i}$. In contrast, the variables $C_{i}$ and $S_{i}$ arise from interference effects and therefore cannot be directly measured in non-interfering $D$ decays.

Next we perform the counting of the number of observables and
parameters to check if we can, in principle, obtain $\gamma$. We
denote by $k$ the number of bins above the $s_{12}=s_{13}$ line so
that there are $2k$ bins in total. We consider $n$ different $B$ decay
modes, such as $B \to D^{(*)}K^{(*)}$. We see that there are $2k$ $C_{i}$'s and $2k$ $S_{i}$'s, 
$n$ of $\delta_{B}^f$ and $r_{B}^f$, and $\gamma$. We end up with
$4k+2n+1$ unknowns and $4kn$ equations. We see that for $n\ge 2$ we
can find $k$ such that there are more observables than unknowns and
thus $\gamma$ can be determined without any approximations. 

We can also determine $\gamma$ in the $n=1$ case using approximations. One
approximation is to use a model for the Dalitz plot.
Another approximation that can be made is to take advantage of the fact that the main correction to $\gamma$ comes from the term that is not proportional to $r_B$ [see, Eq.~(\ref{mainGLW})]. Thus,
dropping terms of order $r_{B}|\epsilon|$ is equivalent to using 
$C_{i}=C_{\bar{i}}$ and $S_{i}=-S_{\bar{i}}$.
In this case we have only $2k+3$ unknowns with $4k$ equations which is solvable for $k\geq2$. 

Some other possibilities to reduce the number of unknowns were discussed
in \cite{Giri:2003ty,Gronau:2001nr,Soffer:1998un,Soffer:1999dz,Atwood:2003mj}. They are applicable here except that one must be careful to distinguish between CP conjugate quantities.

\subsection{Including the kaon time dependence}         
When we outlined how methods for determining $\gamma$ from $B^{\pm}\to DK^{\pm}$ can be adapted to include CP violation in the $\K-\Kbar$ system we glossed over the time dependence of the kaon decay and oscillations by assuming we integrate over all of time. This is not realistic since time-dependent efficiencies must be taken into account. Below we discuss the time dependence issue in more detail.

We begin by writing the time dependence explicitly
\begin{equation}
|\Acal|^{2}=|A_{\pi\pi}|^{2}\left[|A_{S}|^{2}e^{-\Gamma_{S}t}+2e^{-\Gamma
    t}\real\left(\epsilon^{*}A_{L}^{*}A_{S}e^{ix\Gamma
      t}\right)\right],
\end{equation}
\begin{equation}
|\bar{\Acal}|^{2}=|A_{\pi\pi}|^{2}\left[|\bar{A}_{S}|^{2}e^{-\Gamma_{S}t}+2e^{-\Gamma
    t}\real\left(\epsilon^{*}\bar{A}_{L}^{*}\bar{A}_{S}e^{ix\Gamma
      t}\right)\right],
\end{equation}
\begin{equation}
\real(\Acal^{*}\bar{\Acal})=\real(A_{S}^{*}\bar{A}_{S})e^{-\Gamma_{S}t}+\real\left[\epsilon^{*}(A_{S}\bar{A}_{L}^{*}+A_{L}^{*}\bar{A}_{S})e^{\Gamma t(ix-1)}\right],
\end{equation}
\begin{equation}
\imag(\Acal^{*}\bar{\Acal})=\imag(A_{S}^{*}\bar{A}_{S})e^{-\Gamma_{S}t}+\imag\left[\epsilon^{*}(A_{L}^{*}\bar{A}_{S}-A_{S}\bar{A}_{L}^{*})e^{\Gamma t(ix-1)}\right],
\end{equation}
where we have neglected the $|A_{L}|^{2}$ term as it is proportional to $|\epsilon|^{2}\sim10^{-6}$.
In order to translate measurements performed in different timing windows, one must determine the various momentum dependent coefficients of the different time-dependent functions. 
For instance, we will need
\begin{equation}
\int dt\:F(t)|\Acal(t)|^{2},\qquad
\int dt\:F(t)\real(\Acal^{*}\bar{\Acal}),\qquad
\int dt\:F(t)\imag(\Acal^{*}\bar{\Acal}).
\end{equation}	 
where $F(t)$ is the time-dependent detection efficiency defined
above. To obtain such integrals we must determine both the real and imaginary parts of
\begin{equation}
|A_{S}|^{2},\quad A_{L}^{*}A_{S}, \quad A_{S}^{*}\bar{A}_{S},\quad A_{S}^{*}\bar{A}_{L},\quad A_{L}^{*}\bar{A}_{S}
\end{equation} 
for each bin by distinguishing between the $\exp({-\Gamma_{S}t})$, $\exp({-\Gamma t})\cos(x\Gamma t)$, and $\exp({-\Gamma t})\sin(x\Gamma t)$ terms.  When dropping order $r_{B}|\epsilon|$ terms only the first three of these need to be determined. (Some explicit expressions of time integrals can be found in Appendix A.)  

While in principle knowing $F(t)$ will enable us to determine $\gamma$, care must
be taken, as the whole program involves data from different
experiments. For example, one uses charm decay data as an input to the
$B$ decay analysis. In principle $F(t)$ can be very different in such
experiments, so naive use of this data introduces error. In
practice, we expect $F(t)$ to be similar for different experiments so that the induced error will be small. Ultimately estimating this effect depends on the details of particular experiments, so this cannot be done in a general way.

\subsection{Order $r_{B}|\epsilon|$ Terms}
As we have already alluded to, there are complications that arise when
terms of order $r_{B}|\epsilon|$ are included.  This is because there
is a new variable (roughly speaking it is the phase of
$A_{S}^{*}\bar{A}_{S}$) which only appears in interference between
$\D$ and $\Dbar$ amplitudes. To understand how these terms complicate
the analysis, we should introduce new definitions. First, we define
\begin{equation}
A_{D}\equiv A(\D\to\Kbar\pi^{+}\pi^{-}),\qquad
A_{D}r_{D}e^{i\delta_{D}}\equiv A(\D\to\K\pi^{+}\pi^{-}),
\end{equation}
and the CP conjugate amplitudes
\begin{equation}
\bar{A}_{D}\equiv A(\Dbar\to\K\pi^{-}\pi^{+}),\qquad
\bar{A}_{D}\bar{r}_{D}e^{i\bar{\delta}_{D}}\equiv A(\Dbar\to\Kbar\pi^{-}\pi^{+}).
\end{equation}
We emphasize that $A_{D}$, $r_{D}$, $\delta_{D}$ and their conjugates
depend on $(s_{12},s_{13})$ for multi-body decays.  Here $r_{D}$ is a
CKM suppression factor, but we do not know how small it is in any particular region of phase space. From here on we neglect CP violation in the $D$ decay, and then we have
\begin{equation}
A_{D}(s_{12},s_{13})=\bar{A}_{D}(s_{13},s_{12}), \quad
r_{D}(s_{12},s_{13})=\bar{r}_{D}(s_{13},s_{12}), \quad
\delta_{D}(s_{12},s_{13})=\bar{\delta}_{D}(s_{13},s_{12}).
\end{equation}
(In principle $A_{D},r_{D},\delta_{D}$ are each measurable through
flavor specific decays of the kaon, for example, $\D\to(\pi
\ell^+\nu)_{K}\pi^{+}\pi^{-}$ and its CP conjugates.) From these we find
\begin{equation}
A_{S}=\frac{A_{D}}{2pq}\left(qr_{D}e^{i\delta_{D}}-p\right)=\frac{A_{D}}{\sqrt{2}}\left((1-\epsilon)r_{D}e^{i\delta_{D}}-1-\epsilon\right)+O(|\epsilon|^{2}),
\end{equation}
\begin{equation}
A_{L}=\frac{A_{D}}{2pq}\left(qr_{D}e^{i\delta_{D}}+p\right)=\frac{A_{D}}{\sqrt{2}}\left((1-\epsilon)r_{D}e^{i\delta_{D}}+1+\epsilon\right)+O(|\epsilon|^{2}),
\end{equation}
The CP conjugate expressions for $\bar{A}_{S,L}$ are obtained by
$A_{D}\to\bar{A}_{D}$,  $\epsilon\to-\epsilon$, and $A_{S}\to-A_{S}$.
The phase between $p$ and $q$ is unphysical but we adopt the
convention $q =(1-\epsilon)/\sqrt{2}$ and $p=(1+\epsilon)/\sqrt{2}$. The new phase which occurs only in the $r_{B}$ terms which cannot be obtained through measurements of non-interfering $D$ decays alone is that of $A_{D}^{*}\bar{A}_{D}$ so we define
\begin{equation}
\theta_{D}(s_{12},s_{13})\equiv\arg(A_{D}^{*}\bar{A}_{D}).
\end{equation}
Under an interchange of the pion momenta one has $\theta_{D}\to-\theta_{D}$.  The value of this angle at a point in phase space is unphysical, but for a particular choice of $\delta_{B}$ it is fixed, and its momentum dependence is physical.  It reduces to $\delta_{12,13}-\delta_{13,12}$ of \cite{Giri:2003ty} in the $\epsilon\to0$ limit.  

If one expands out the real and imaginary parts of $\Acal^{*}\bar{\Acal}$ one will obtain various combinations of $|A_{D}|^{2}, r_{D},\delta_{D}$ and trig functions of $\theta_{D},\delta_{D},\bar{\delta}_{D}$ such as
$r_{D}\cos(\theta-\delta_{D})$,
$\bar{r}_{D}\cos(\theta+\bar{\delta}_{D})$, and
$r_{D}\bar{r}_{D}\cos(\theta+\bar{\delta}_{D}-\delta_{D})$.
These are of course multiplied by $|A_{D}\bar{A}_{D}|$ and are ultimately integrated over bins of phase space.  The full result can be found in Appendix A.


\section{Assuming Breit-Wigner Dependence}
As we alluded to previously, some unknowns can be eliminated by
assuming a Breit-Wigner dependence for the $D$ decays. The
change one needs to keep in mind when generalizing the 
discussion of Breit-Wigner dependence from a case where kaon CP
violation is neglected is that CP
conjugate amplitudes are no longer related by only an exchange of pion
momenta. 

We substitute $A_{S}$ above with a sum over Breit-Wigner functions 
\begin{equation}
A_{S}(s_{12},s_{13})=a_{0}e^{i\delta_{0}}+\sum_{r}a_{r}e^{i\delta_{r}}A_{r}(s_{12},s_{13}),
\label{BWdecomp}
\end{equation}
where
\begin{equation}
A_{r}(s_{12},s_{13})={^{J}\!\mathcal{M}}_{r}W^{r},
\end{equation}
and the index $r$ labels the resonance.  The factor ${^{J}\!\mathcal{M}}_{r}$ depends on the spin of the resonance, for example ${^{0}\mathcal{M}}_{r}=1$, ${^{1}\mathcal{M}}_{r}=-2\vec{k}_{1}\cdot\vec{k}_{2}$ where $\vec{k}_{1},\vec{k}_{2}$ are the spatial momenta of the particles originating from the resonance.  The Breit-Wigner function is
\begin{equation}
W^{r}=\frac{1}{s-M_{r}^{2}+iM_{r}\Gamma_{r}(\sqrt{s})},
\end{equation}
where $M_{r}$ is the mass of the resonance.  The argument of $W^{r}$
depends on the particles which participate in the resonance, for
example for the $\rho^{0}$ the argument of $W^{r}$ is $s_{23}$.
Explicit expressions for the mass dependent width
$\Gamma_{r}(\sqrt{s})$ and the other ${^{J}\!\mathcal{M}}_{r}$'s can be
found in \cite{ResStuff,delAmoSanchez:2010rq,Poluektov:2010wz,Muramatsu:2002jp}.

In order to account for CP violation in the kaon system one cannot
assume that the amplitudes are related to their CP conjugates by a
momentum exchange alone, for instance
\begin{equation}
a_{\rho}\propto A(\D\to\rho^{0}K_{S})\neq \bar{a}_{\rho} \propto A(\Dbar\to\rho^{0}K_{S}).
\end{equation}
The reason for this is that the $K_{S}$ is not an exactly even superposition of $\K$ and $\Kbar$.  Fortunately, these are related through simple momentum independent factors
\begin{equation}
A(\D\to\rho^{0}K_{S})=\frac{A_{\rho}}{2pq}(qR_{\rho}-p)=\frac{A_{\rho}}{\sqrt{2}}\left(R_{\rho}-\epsilon R_{\rho}-1-\epsilon\right) +O(|\epsilon|^{2}),
\end{equation}
\begin{equation}
A(\Dbar\to\rho^{0}K_{S})=\frac{A_{\rho}}{2pq}\left(q-pR_{\rho}\right)=\frac{A_{\rho}}{\sqrt{2}}\left(1-\epsilon-R_{\rho}-\epsilon R_{\rho}\right) +O(|\epsilon|^{2}),
\end{equation}
where $A_{\rho}\equiv A(\D\to\rho^{0}\Kbar)$, $A_{\rho}R_{\rho}\equiv
A(\D\to\rho^{0}\K)$ with $R_{\rho}$ complex and we have neglected CP violation in the $D$ decay so that $A(\D\to\rho^{0}\Kbar)=A(\Dbar\to\rho^{0}\K)$.  Similar expressions apply for any resonance decay of the form $\D\to K_{S}X$.

The situation is slightly different for decays in which the $K_{S}$
emerges from a resonance such as $\D\to K^{*-}\pi^{+}$.  In this case 
\beq
a_{K^{*}}\propto A(\D\to K^{*-}\pi^{+})\ne \bar{a}_{K^{*}}\propto
A(\Dbar\to K^{*+}\pi^{-})
\eeq 
because $A(K^{*-}\to K_{S}\pi^{-})\neq A(K^{*+}\to K_{S}\pi^{+})$.  Fortunately, these are again simply related by momentum-independent factors
\begin{equation}
A(K^{*-}\to K_{S}\pi^{-})=-\frac{A_{*}}{2q}=-\frac{A_{*}}{\sqrt{2}}(1+\epsilon),
\end{equation}
\begin{equation}
A(K^{*+}\to K_{S}\pi^{+})=\frac{A_{*}}{2p}=\frac{A_{*}}{\sqrt{2}}(1-\epsilon),
\end{equation}
where $A_{*}=A(K^{*-}\to\Kbar\pi^{-})=A(K^{*+}\to\K\pi^{+})$ and we have assumed $\Delta(\mbox{strangeness})=2$ decays to be forbidden entirely.  In this way one can relate $a_{K^{*}}$ to $\bar{a}_{K^{*}}$.

For $\epsilon\to0$ one only has the amplitude $A_{S}$, but for finite
$\epsilon$, we also have $A_{L}$, which can be decomposed just like (\ref{BWdecomp}).  The same discussion on how to account for CP violation applies, however the relation between some of the coefficients is different.  A general procedure for relating amplitude coefficients is as follows
\begin{enumerate}
\item Express decay amplitudes in terms of $A(D\to KX)$ for resonances which do not decay to the final state neutral kaon, where $K$ is one of $\K,\Kbar$.

\item Express decay amplitudes in terms of $A(X_{s}\to KX)$ where $X_{s}$ is a resonance which decays to a neutral kaon.

\item Project out the $K_{S,L}$ component using the reciprocal basis
\begin{equation}
\bra{K_{L,S}}=\frac{1}{p}\bra{\K}\pm\frac{1}{q}\bra{\Kbar}
\end{equation}
to relate the various coefficients.
\end{enumerate}
where the $+$ is for the $K_{L}$ and the $-$ is for the $K_{S}$.

\section{Estimating the error in $\gamma$}

Next we estimate the error introduced on the extracted value of
$\gamma$ if one were to use the standard analysis and neglect CP violation in the kaon system. We do so by making the following simplifying assumptions: We neglect terms of order $r_{B}|\epsilon|$, we assume that we integrate over all of the $K_{S}\pi^{+}\pi^{-}$ phase space, and we assume that $F(t)=1$.  A simple expression for $\gamma$ can be obtained in terms of the difference
\begin{equation}
D_{0\infty}=\int_{0}^{\infty}dt\left[\Gamma(B^{+}\to(\pi^{4})_{D}K^{+})-\Gamma(B^{-}\to(\pi^{4})_{D}K^{-})\right].
\end{equation}
Inverting the result for $D_{0\infty}$ one finds
\begin{equation}
\sin(\gamma)=\frac{D_{0\infty}/N-4\kappa'_{D}\real(\epsilon)}{4r_{B}F'_{D}\sin(\delta_{B})}.
\end{equation}
where $N\equiv|A(K_{S}\to\pi\pi)A_{B}|^{2}/2\Gamma_{S}$ and $F'_{D}$ and $\kappa'_{D}$ are functions of $A_{D},r_{D},\delta_{D}$ and $\bar{A}_{D},\bar{r}_{D},\bar{\delta}_{D}$ which are defined in Appendix A.
Next we define
\begin{equation}
\Delta\gamma\equiv\gamma-\gamma_{0},
\end{equation}
where $\gamma_{0}$ is $\gamma$ as determined from $A_{CP}$ with $\epsilon=0$.  We find
\begin{equation}
\Delta\gamma=-\frac{\kappa'_{D}\real(\epsilon)}{r_{B}F'_{D}\cos(\gamma_{0})\sin(\delta_{B})},
\label{Deltagamma}
\end{equation}
and we expect $\kappa'_{D}/F'_{D}\sim1$.  To see this, note that in the limit where we neglect doubly Cabbibo suppressed $D$ decays and let the $D$ decay to a true CP eigenstate one obtains the same expression for $\Delta\gamma$ with $\kappa'_{D}/F'_{D}=1$.
Therefore we find $\Delta\gamma\sim|\epsilon|/r_{B}$.


\begin{figure}
\begin{center}
\epsfig{file=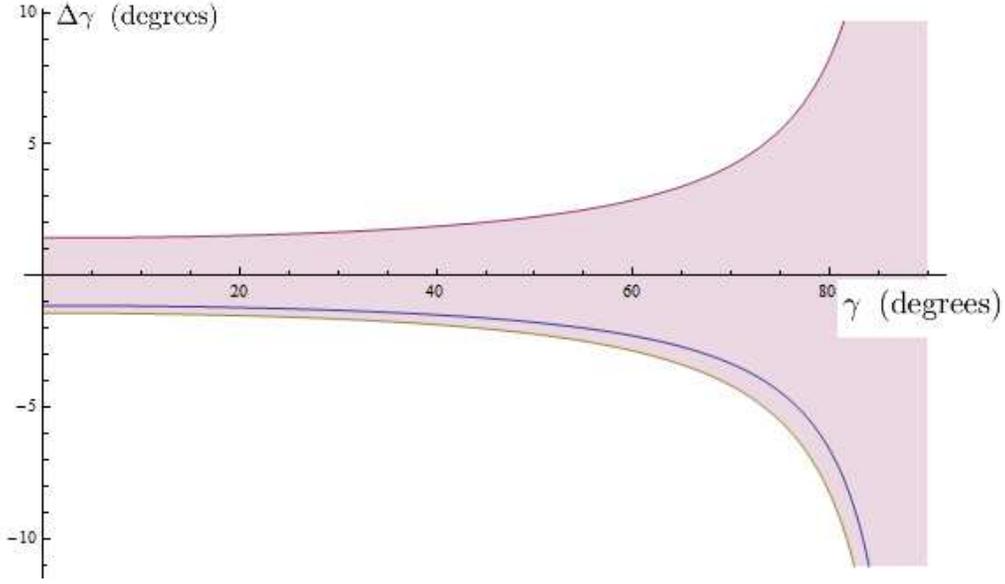, height=8.0cm}
\caption{\footnotesize{ $\Delta\gamma$ (blue line) as a function of $\gamma$, as it
  appears in (\ref{Deltagamma}) for
  $\kappa'_{D}/F'_{D}=1,r_{B}=10^{-1},\delta_{B}=\pi/2$.  The shaded
  region represents an error due to $\delta(D_{0\infty}/N)=0.01$.  For
  kaon mixing to be relevant, uncertainty in the CP asymmetry
  $D_{0\infty}$ must be small enough that the blue line lies outside
  the shaded region.}} \label{fig:deltagamma}
\end{center}
\end{figure}

There appear to be several limits in which $\Delta\gamma$ diverges.
The first type is when any of $r_{B}$, $F'_{D}$, or $\delta_{B}$
vanish.  None of these are problematic since they arise only when working to leading order, which is no longer justified in that case. Keeping the
full expression we find
\begin{equation}
\frac{\sin(\gamma_{0}+\Delta\gamma)-\sin(\gamma_{0})}{\sin(\gamma_{0})}=-\frac{4\kappa'_{D}\real(\epsilon)}{D_{0\infty}/N},
\end{equation}
which does not depend on $r_{B}$, $F'_{D}$, or $\delta_{B}$. 
Next we see that $\Delta\gamma$ appears to diverge for
$\gamma_{0}\to\pi/2$.  This divergence reflects the fact that
$D_{0\infty}$ depends only very weakly on $\gamma$ for
$\gamma\approx\pi/2$.  Therefore, any source of error in
$D_{0\infty}/N$ would also cause a large shift in $\gamma$ in this
region.  The uncertainty in $\gamma$ using this method is then large
for $\gamma\approx\pi/2$, so we expect that $\Delta\gamma$ is also
large there.  This uncertainty is only intrinsic to the Dalitz method
where $S_{i}$ is small.  Where $C_{i}$ is small there is a similar
uncertainty near $\gamma\approx0$. The effect is demonstrated in Fig.~\ref{fig:deltagamma}.

The best current determinations of $\gamma$ from Belle and BaBar have uncertainties of roughly $\pm10^{o}$ \cite{Poluektov:2010wz,delAmoSanchez:2010rq}.  We expect $|\Delta\gamma|$ to be of order $\real(\epsilon)/r_{B}\approx1^{o}$ so it may be some time before the effect of CP violation in kaon mixing and decay becomes relevant.  This correction is, however, very large compared to the largest irreducible theoretical error on the determination of $\gamma$ from $B\to DK$ \cite{Brod:2013sga}.

\section{Conclusion}

The $B \to DK$ program is known to have the smallest theoretical error
in any determination of weak parameters \cite{Brod:2013sga}. As precision improves it becomes more important to look for sub-leading effects \cite{Schlupp:2013vfa,Aushev:2010bq}. 
We generalized the $B \to DK$ method for determining $\gamma$ to
account for additional sources of CP violation from mixing of final
state neutral kaons. We found that $\gamma$ is shifted by an amount of
order $\epsilon/r_B \sim 10^{-2}$. While this effect is still below the
current experimental sensitivity, it may be important in the near
future. We have discussed how existing methods can be corrected to account for this effect.

\acknowledgments{
We thank Abner Soffer For useful discussions.
YG is a Weston Visiting Professor at the Weizmann Institute.
This work was partially supported by a grant from the Simons Foundation ($\#$267432 to Yuval Grossman).
The work of YG is supported in part by the U.S. National
Science Foundation through grant PHY-0757868 and
by the United States-Israel Binational Science
Foundation (BSF) under grant No.~2010221.}

\appendix
\section{Some Explicit Expressions}

In this appendix we collect more general expressions that we omit in
the main text.  In the main text we usually kept the $\gamma$ dependence explicit, while here we keep the time dependence explicit.  Consider a $D$ meson state which results from the decay of the $B^{-}$ and which is a superposition of $\D$ and $\Dbar$ depending on $A_{B},r_{B},\delta_{B},\gamma$.  We denote that state $D_{B}$.  It is useful to write quantities in terms of $D_{B}$ decay amplitudes.  We define
\begin{equation}
c_{\K}\equiv A(D_{B}\to\K\pi^{+}\pi^{-}),\qquad
c_{\Kbar}\equiv A(D_{B}\to\Kbar\pi^{+}\pi^{-}).
\end{equation}
We also define $c_{L,S}$ to be $c_{\K}$ with the $\K$ replaced by $K_{L,S}$.  These amplitudes are simply related
\begin{equation}
c_{S}\equiv\frac{1}{2pq}\left(qc_{\K}-pc_{\Kbar}\right), \qquad
c_{L}\equiv\frac{1}{2pq}\left(qc_{\K}+pc_{\Kbar}\right),
\end{equation}
and can be expressed in terms of $\gamma$
\begin{equation}
c_{\K}=A_{B}\left(A_{D}r_{D}e^{i\delta_{D}}+\bar{A}_{D}r_{B}e^{i(\delta_{B}-\gamma)}\right),
\end{equation}
\begin{equation}
c_{\Kbar}=A_{B}\left(A_{D}+\bar{A}_{D}r_{B}\bar{r}_{D}e^{i(\delta_{B}+\bar{\delta}_{D}-\gamma)}\right).
\end{equation}
These can be written for the $B^{+}$ by $\gamma\to-\gamma$ and
$(A_{D},r_{D},\delta_{D})\leftrightarrow(\bar{A}_{D},\bar{r}_{D},\bar{\delta}_{D})$.
The latter is equivalent to changing the sign of $\epsilon$ in the
overall decay width.  The kaon time dependent overall width of the $B \to DK$ followed by
$D \to K \pi \pi$ and then $K \to \pi\pi$
\begin{equation}
\frac{d^{2}\Gamma}{ds^{2}}(B^{-}\to f^-)=|A_{\pi\pi}|^{2}\left(|c_{S}|^{2}e^{-\Gamma_{S}t}+|\epsilon c_{L}|^{2}e^{-\Gamma_{L}t}+2e^{-\Gamma t}\real(\epsilon^{*}c_{L}^{*}c_{S}e^{ix\Gamma t})\right),
\end{equation}
where $f^-$ is defined in (\ref{eq:deff}).
When expanding $|c_{S}|^{2}$ and $c_{L}^{*}c_{S}$ there are a number of expressions which occur frequently, we define
\beqa
F_{D}&\equiv&|A_{D}|^{2}\left(1+r_{D}^{2}-2r_{D}\cos(\delta_{D})\right),
\nonumber \\
F'_{D}&\equiv&\real(A_{D}^{*}\bar{A}_{D})-2r_{D}\real\left(A_{D}^{*}\bar{A}_{D}e^{-i\delta_{D}}\right)-r_{D}\bar{r}_{D}\real\left(A_{D}^{*}\bar{A}_{D}e^{i(\bar{\delta}_{D}-\delta_{D})}\right),
\nonumber \\
\kappa_{D}&\equiv&|A_{D}|^{2}\left(1-r_{D}^{2}-2r_{D}\sin(\delta_{D})\right),
\nonumber \\
\kappa'_{D}&\equiv&|A_{D}|^{2}\left(1-r_{D}^{2}+2r_{D}\sin(\delta_{D})\right),
\eeqa
with barred quantities can be obtained by
$(A_{D},r_{D},\delta_{D})\to(\bar{A}_{D},\bar{r}_{D},\bar{\delta}_{D})$. Neglecting
terms of order $r_{B}|\epsilon|$ we get, for example,
\begin{equation}
|c_{S}|^{2}=\frac{|A_{B}|^{2}}{2}\left[F_{D}+r_{B}^{2}\bar{F}_{D}+2\kappa_{D}\real(\epsilon)-2r_{B}F'_{D}\cos(\delta_{B}-\gamma)\right],
\end{equation}
\begin{equation}
\epsilon^{*}c_{L}^{*}c_{S}=-\frac{|A_{B}A_{D}|^{2}}{2}\epsilon^{*}\left(1-r_{D}^{2}-2ir_{D}\sin(\delta_{D})\right).
\end{equation}
When integrating over all of time, these combine giving expressions which include $\kappa'_{D}$, so that $\kappa'_{D}/F'_{D}$ appears in the expression for $\Delta\gamma$.

Integrals over time involve a detection efficiency function $F(t)$
which must be determined for a particular experiment.  As an
illustrative example, consider $F(t)=1$ for $t_{1}<t<t_{2}$
and $F(t)=0$ otherwise.  The following integrals involving an
arbitrary complex number $z$ may also be useful
\begin{equation}
\Gamma_{S}\int_{t_{1}}^{t_{2}}dt\:\real(z^{*}e^{\Gamma t(ix-1)})=(x\real(z)-\imag(z))Q_{c}(t_{1},t_{2})+(\imag(z)-x\real(z))Q_{s}(t_{1},t_{2}),
\end{equation}
\begin{equation}
\Gamma_{S}\int_{t_{1}}^{t_{2}}dt\:\imag(z^{*}e^{\Gamma t(ix-1)})=(x\real(z)-\imag(z))Q_{c}(t_{1},t_{2})+(\real(z)+x\imag(z))Q_{s}(t_{1},t_{2}),
\end{equation}
where
\begin{equation}
Q_{c}(t_{1},t_{2})=-\frac{1-y}{1+x^{2}}\left[e^{-\Gamma t_{2}}\cos(x\Gamma t_{2})-e^{-\Gamma t_{1}}\cos(x\Gamma t_{1})\right],
\end{equation}
\begin{equation}
Q_{c}(t_{1},t_{2})=-\frac{1-y}{1+x^{2}}\left[e^{-\Gamma t_{2}}\sin(x\Gamma t_{2})-e^{-\Gamma t_{1}}\sin(x\Gamma t_{1})\right],
\end{equation}
where we have used $\Gamma_{S}=(1-y)\Gamma$. While we have not used
any approximation involving the variables $x,y$, it greatly simplifies things to use $x\approx -y\approx1$, which we use in the main body of the paper when estimating $\Delta\gamma$.  When integrating over all time we have $Q_{c}(0,\infty)=(1-y)/(1+x^{2})\approx1$ and $Q_{s}(0,\infty)=0$.

It is also useful to have an expansion of $\Acal^{*}\Acal$.  Note that  
\begin{equation}
\real(\Acal^{*}\Acal)=\real(A_{S}^{*}\bar{A}_{S})e^{-\Gamma_{S}t}+|\epsilon|^{2}\real(A_{L}^{*}\bar{A}_{L})e^{-\Gamma_{L}t}+\real\left[\epsilon^{*}(A_{S}\bar{A}_{L}^{*}+A_{L}^{*}\bar{A}_{S})e^{\Gamma t(ix-1)}\right],
\end{equation}
\begin{equation}
\imag(\Acal^{*}\Acal)=\imag(A_{S}^{*}\bar{A}_{S})e^{-\Gamma_{S}t}+|\epsilon|^{2}\imag(A_{L}^{*}\bar{A}_{L})e^{-\Gamma_{L}t}+\imag\left[\epsilon^{*}(-A_{S}\bar{A}_{L}^{*}+A_{L}^{*}\bar{A}_{S})e^{\Gamma t(ix-1)}\right].
\end{equation}
Therefore, in terms of
$\theta_{D}(s_{12},s_{13})\equiv\arg(A_{D}^{*}\bar{A}_{D})$ to order
$|\epsilon|$ we obtain
\begin{align}
A_{S}^{*}\bar{A}_{S}&=\frac{|A_{D}^{*}\bar{A}_{D}|}{2}e^{i\theta_{D}}\left[1+(1-2\real(\epsilon))r_{D}\bar{r}_{D}e^{i(\bar{\delta}_{D}-\delta_{D})}-r_{D}e^{-i\delta_{D}}-\bar{r}_{D}e^{i\bar{\delta}_{D}}\right. \\ 
& \hspace*{35mm}\left.-2ir_{D}e^{-i\delta_{D}}\imag(\epsilon)+2i\bar{r}_{D}e^{i\bar{\delta}_{D}}\imag(\epsilon)\right], \\
\epsilon^{*} A_{S}\bar{A}_{L}^{*}&=\epsilon^{*}\frac{|A_{D}^{*}\bar{A}_{D}|}{2}e^{-i\theta_{D}}\left(r_{D}e^{i\delta_{D}}-1-\bar{r}_{D}e^{-i\bar{\delta}_{D}}+r_{D}\bar{r}_{D}e^{i(\delta_{D}-\bar{\delta}_{D})}\right).
\end{align}

\section{Other Small Weak Phases}

There are two other sources of CP violation which we have not taken into account, but which may become important at the $r_{B}|\epsilon|$ level.  Neither of these present any significant complication to the methods discussed in the body of the letter where they are largely ignored, and they are simply accounted for, at least to the extent to which they can be accurately measured in other experiments.

\subsection{In $D\to K\pi^{+}\pi^{-}$}
There is a small amount of CP violation involved in the decay of the $D$.  This plays much the same role as the CP violation in the kaon mixing and decay, both are like CP violation in the $D$ decay for the intents and purposes of the Dalitz analysis.  Regardless of how small the CP violation is in the $D$ decay itself, this is already taken into account when neglecting $r_{B}|\epsilon|$ terms when we allow for $T^{+}_{i}\neq T^{-}_{\bar{i}}$.  This CP violation makes its appearance in the variables $\delta_{D}$ and $\bar{\delta}_{D}$.  This can be taken into account for all terms (including $r_{B}|\epsilon|$ terms) by allowing for $\delta_{D}(s_{12},s_{13})\neq\bar{\delta}_{D}(s_{13},s_{12})$.  In fact
\begin{equation}
\frac{\delta_{D}(s_{12},s_{13})-\bar{\delta}_{D}(s_{13},s_{12})}{2}=\arg\left(\frac{V_{us}V_{cd}^{*}}{V_{ud}V_{cs}^{*}}\right).
\end{equation}
Note that $V_{us}V_{cd}^{*}/V_{ud}V_{cs}^{*}\sim\lambda^{2}$ with a non-zero phase coming in at $O(\lambda^{6})$.  This is of course extremely small, but it may be compensated for by kinematic effects in some regions of phase space (i.e. where $r_{D}$ is not small).  In that case the phase can be thought of as a $\lambda^{4}$ correction, which may be competitive with $O(|\epsilon|)$ effects.  As noted however, this is most important in the $T_{i}^{\pm}$ terms where it is already taken into account, otherwise it is suppressed by $r_{B}$.

\subsection{In $B^{\pm}\to DK^{\pm}$}
The weak angle involved in the decay $B^{\pm}\to DK^{\pm}$ is not precisely $\gamma$ but receives some small corrections from other CKM elements.  The ratio of CKM matrix elements involved in the decay is $V_{cs}V_{ub}^{*}/V_{us}V_{cb}^{*}$.  Relating this to $\gamma$ involves
\begin{equation}
\frac{V_{cd}V_{cs}}{V_{ud}V_{us}}=-1+A^{2}\lambda^{4}-A^{2}\lambda^{4}(\rho+i\eta)+O(\lambda^{6}).
\end{equation}
From this we see that corrections to the phase $e^{-i\gamma}$ come in at order $\lambda^{4}\sim10^{-3}$.  This is still unimportant at the $|\epsilon|/r_{B}$ level but is competitive with $O(|\epsilon|)$ corrections.  In all previous sections where we have written $\gamma$, we in fact mean $\gamma+\phi_{B}$ where
\begin{equation}
\phi_{B}\equiv\arg\left(\frac{V_{cd}V_{cs}}{V_{ud}V_{us}}\right).
\end{equation}
This can be considered a fundamental limitation on $B^{\pm}\to DK^{\pm}$ methods, although it can be overcome to whatever accuracy $\phi_{B}$ can be measured by other means.




\begin{thebibliography}{99}



\bibitem{BLS}
  G.~C.~Branco, L.~Lavoura and J.~P.~Silva,
  ``CP violation'', Clarendon Press (1999).

\bibitem{Gronau:1991dp}
  M.~Gronau and D.~Wyler,
  Phys.\ Lett.\ B {\bf 265}, 172 (1991).

\bibitem{Atwood:2000ck} 
  D.~Atwood, I.~Dunietz and A.~Soni,
  Phys.\ Rev.\ D {\bf 63}, 036005 (2001)
  [hep-ph/0008090].

\bibitem{Giri:2003ty} 
  A.~Giri, Y.~Grossman, A.~Soffer and J.~Zupan,
  Phys.\ Rev.\ D {\bf 68}, 054018 (2003)
  [hep-ph/0303187].


\bibitem{Poluektov:2004mf} 
  A.~Poluektov {\it et al.}  [Belle Collaboration],
  Phys.\ Rev.\ D {\bf 70}, 072003 (2004)
  [hep-ex/0406067].
  
\bibitem{Gronau:1990ra}
  M.~Gronau and D.~London,
  Phys.\ Lett.\ B {\bf 253}, 483 (1991).

\bibitem{Kayser:1999bu} 
  B.~Kayser and D.~London,
  Phys.\ Rev.\ D {\bf 61}, 116013 (2000)
  [hep-ph/9909561].
  
\bibitem{Azimov:1980}
  Y.~I.~Azimov and A.A, Iogansen,
  Yad. Fiz. {\bf 33}, 388 (1981)
  [Sov. J. Nucl. Phys. {\bf 33}, 205 (1981)].
  
\bibitem{Azimov:1998sz} 
  Y.~I.~Azimov,
  Eur.\ Phys.\ J.\ A {\bf 4}, 21 (1999)  [hep-ph/9808386].

\bibitem{Lipkin:1999qz}
  H.~J.~Lipkin, Z.~-z.~Xing,
  Phys.\ Lett.\  {\bf B450}, 405-411 (1999).
  [hep-ph/9901329].

\bibitem{Bigi:2005ts}
  I.~I.~Bigi, A.~I.~Sanda,
  Phys.\ Lett.\  {\bf B625}, 47-52 (2005).
  [hep-ph/0506037].

\bibitem{Calderon:2007rg}
  G.~Calderon, D.~Delepine, G.~L.~Castro,
  Phys.\ Rev.\  {\bf D75}, 076001 (2007).
  [hep-ph/0702282 [HEP-PH]].

\bibitem{Grossman:2011zk} 
  Y.~Grossman and Y.~Nir,
  JHEP {\bf 1204}, 002 (2012)
  [arXiv:1110.3790 [hep-ph]].

\bibitem{Grossman:2005rp} 
  Y.~Grossman, A.~Soffer and J.~Zupan,
  Phys.\ Rev.\ D {\bf 72}, 031501 (2005)
  [hep-ph/0505270].


\bibitem{Silva:1999bd} 
  J.~P.~Silva and A.~Soffer,
  Phys.\ Rev.\ D {\bf 61}, 112001 (2000)
  [hep-ph/9912242].

\bibitem{Amorim:1998pi} 
  A.~Amorim, M.~G.~Santos and J.~P.~Silva,
  Phys.\ Rev.\ D {\bf 59}, 056001 (1999)
  [hep-ph/9807364].

\bibitem{Rama:2013voa} 
  M.~Rama,
  arXiv:1307.4384 [hep-ex].

\bibitem{Gronau:2001nr} 
  M.~Gronau, Y.~Grossman and J.~L.~Rosner,
  Phys.\ Lett.\ B {\bf 508}, 37 (2001)
  [hep-ph/0103110].

\bibitem{Soffer:1998un} 
  A.~Soffer,
  hep-ex/9801018.


\bibitem{Soffer:1999dz} 
  A.~Soffer,
  Phys.\ Rev.\ D {\bf 60}, 054032 (1999)
  [hep-ph/9902313].

\bibitem{Atwood:2003mj} 
  D.~Atwood and A.~Soni,
  Phys.\ Rev.\ D {\bf 68}, 033003 (2003)
  [hep-ph/0304085].
  
\bibitem{ResStuff}
  E.~M.~Aitala {\it et al.} [E791 Collaboration], 
  Phys.\ Rev.\ Lett.\ {\bf86}, 770 (2001).

\bibitem{delAmoSanchez:2010rq} 
  P.~del Amo Sanchez {\it et al.}  [BaBar Collaboration],
  Phys.\ Rev.\ Lett.\  {\bf 105}, 121801 (2010)
  [arXiv:1005.1096 [hep-ex]].

\bibitem{Poluektov:2010wz} 
  A.~Poluektov {\it et al.}  [Belle Collaboration],
  Phys.\ Rev.\ D {\bf 81}, 112002 (2010)
  [arXiv:1003.3360 [hep-ex]].

\bibitem{Muramatsu:2002jp} 
  H.~Muramatsu {\it et al.}  [CLEO Collaboration],
  Phys.\ Rev.\ Lett.\  {\bf 89}, 251802 (2002)
  [Erratum-ibid.\  {\bf 90}, 059901 (2003)]
  [hep-ex/0207067].

  
\bibitem{Brod:2013sga} 
  J.~Brod and J.~Zupan,
  arXiv:1308.5663 [hep-ph].
  
\bibitem{Schlupp:2013vfa} 
  M.~Schlupp,
  arXiv:1307.7018 [hep-ex].

\bibitem{Aushev:2010bq} 
  T.~Aushev, W.~Bartel, A.~Bondar, J.~Brodzicka, T.~E.~Browder, P.~Chang, Y.~Chao and K.~F.~Chen {\it et al.},
  arXiv:1002.5012 [hep-ex].







 













\end{thebibliography}
\end{document}